# Abundance gradients in the galactic disk

CUI Chenzhou (崔辰州), CHEN Yuqin (陈玉琴), ZHAO Gang (赵　刚)
& ZHAO Yongheng (赵永恒)

National Astronomical Observatories, Chinese Academy of Sciences, Beijing 100012, China
Correspondence should be addressed to Cui Chenzhou (email: ccz@bao.ac.cn)

**Abstract** The relationship between abundances and orbital parameters for 235 F- and G-type intermediate- and low- mass stars in the Galaxy is analyzed. We found that there are abundance gradients in the thin disk in both radial and vertical directions (–0.116 dex $kpc^{-1}$ and –0.309 dex $kpc^{-1}$ respectively). The gradients appear to be flatter as the Galaxy evolves. No gradient is found in the thick disk based on 18 thick disk stars. These results indicate that the ELS model is mainly suitable for the evolution of the thin disk, while the SZ model is more suitable for the evolution of the thick disk. Additionally, these results indicate that in-fall and out-flow processes play important roles in the chemical evolution of the Galaxy.

The formation and evolution of the Galaxy has been a hot topic for many years. In terms of the galactic formation, the ELS model[1] and the SZ model[2] are the two most probable scenarios. According to the ELS model, the Galaxy is formed through the collapse of a single-prototype galactic cloud. In the placid and rapid collapse process, the high-dispersion, spherical, slowly-rotating, and metal-poor halo turns into a high-density, flat, rapidly-rotating, and metal-rich disk. At the same time, supernovae explosions continually enrich the interstellar medium (ISM). As a result, obvious abundance gradients emerge. From the galactic halo to the disk, the abundances become higher. According to the picture described by the SZ model, the Galaxy is not formed from the collapse of a single-prototype galactic cloud, but from inter-collision or accretion of multiple-prototype galactic clouds that have different evolution histories. As suggested by the SZ model, the mixing of many components erases correlations among age, abundance and kinematics.

The abundance gradient in the galactic disk is of great importance in understanding the formation and evolution of the disk. The variance of the gradients on a space and time scale reflects not only the enrichment history of the ISM but also the in-fall and out-flow processes of the Galaxy, and thus is an important constraint for models of galactic chemical evolution. According to the scenario that the Galaxy is formed from inside to outside, the inner disk has higher medium density, higher star formation rate (SFR) and higher chemical evolution rate. Therefore, the abundances in the inner disk region are higher than those in the outer region, i.e. the re is an abundance gradient. But, some processes occur that affect galactic chemical evolution. For example, galactic dynamical evolution, the movements of stars, and the interactions among stars and giant molecular clouds, will affect the distribution of chemical elements, even erase the abundance gradients. Whether abundance gradients of chemical elements, and their variances in space and time exist in the Galaxy or in the galactic disk is directly related to the mechanism of galactic formation and dynamical processes that occur in the Galaxy, and are important constraints for chemical evolution models for the Galaxy.

Observationally, many types of objects are tracers of galactic abundance gradients. On one hand, H II regions and early-type main sequence stars, especially B-type stars, are often used to study the abundance gradients of the present Galaxy. These two types of objects have just been formed from the present galactic ISM and are still close to the places where they formed. To some extent, their abundances are a representation of the chemical composition of the present ISM at these objects' birthplaces. On the other hand, planetary nebulas (PNs) are formed from surface materials that have erupted from former stars; abundances of PNs represent the chemical composition of the ancient ISM from which their former stars were formed, and are indicators for the ancient abundance gradients. At the same time, intermediate- and old-age open clusters and late-type main sequence stars have very large age span, and can be used to trace abundance gradient variances in space and time scales.

Maciel[3], Hou and Chang[4] presented general reviews for the study of galactic abundance gradients. Up to now, many of the observations and statistical analysis that have been carried out suggest that an abundance gradient of $-0.07$ dex kpc$^{-1}$ exists in the galactic disk. At least, the research results measuring the abundances of oxygen in young objects are coincident with it on a large scale. However, both quantitively and qualitatively, the research on abundance gradients has not gotten the last word: different research, different samples, and different analysis methods usually provide quite different results. For example, for the abundance observation of B-type stars, Fitzsimmons et al. [5] do not find any obvious radial abundance gradient; Kaufer et al.[6] also point out that there is no abundance gradient, at least in the region of galactocentric distance $R > 6$ kpc; but, Smartt and Rolleston[7] consider that in the region of $R$ between 6 and 18 kpc, abundance gradient exists; Rolleston[8] also draws the conclusion that an abundance gradient exists in the galactic disk. The reasons for these completely different results may be the difference in selected samples, error in abundance analysis, error in orbit integration, and so on. Presently, it is impossible to eliminate the interference of these factors. A possible solution is to verify different conclusions independently with other samples. Especially important are samples on which high-resolution spectrum observation can be carried out to obtain reliable kinematic orbits, which have crucial significance in clarifying the problem. In this paper, we described the detailed statistical analysis carried out on the abundance gradients in the galactic disk, using two sets of high-resolution observation data for F and G type stars provided by Chen et al.[9] and Edvardsson et al.[10]. The corresponding orbital parameters were obtained from the stellar orbit program provided by Cui et al.[11] To some extent, this paper describes subsequent research from our previous work[11] (hereafter Cui2000). But compared with Cui2000, sample data in this paper has stricter selection criteria.

Abundance measurement has better coherence and is more precise. Major improvements are as follows:

(i) All samples are F and G late-type, intermediate- and low-mass stars. Production of inner nuclear reactions hardly affects their surface chemical composition, which represents the abundances in the ISM when and where they were formed. Intermediate- and low-mass stars have very large age spans; their age information can be used to study the variances in abundance gradients on a time scale.

(ii) Stellar abundances are obtained using high-resolution spectrums and an improved analysis method; thus the accuracy of these data is very high.

(iii) Although the analysis samples are selected from two different research groups, we adopt uniform procedures to calculate abundances, ages and dynamic orbits to ensure the coherence of the data.

(iv) We adopt the mean radius ($R_m$) of stellar orbit as the analysis factor in the radial direction. $R_m$ is a more reasonable representation of stellar orbit characteristics than the maximum galactocentric distance ($R_{max}$)[12,13].

(v) The samples are strictly screened, and stars that belong to different populations or represent unreliable data are excluded.

## 1 Data selection

In this paper, we analyze the abundance gradients of 10 elements, such as Fe, O and Na, for 235 F and G-type stars. These samples are taken from Chen et al. [9] and Edvardsson et al. [10]. Chen et al. carried out high-resolution, high signal-to-noise ratio (S/N) spectrum observation for 90 F and G-type stars with the 2.16 m telescope at National Astronomical Observatories (Xinglong, China), and obtained the abundances of 13 elements (such as Fe, O and Na). Edvardsson et al. obtained high-resolution and high S/N spectrums of 189 F and G-type stars with the 1.4 m telescope at Europe Southern Observatory (La Silla, Chile) and 2.7 m telescope at McDonald Observatory (Davis Mountains, USA).

We re-analyze the observation data from Edvardsson et al.[10] and get the abundances and age using improved procedures:

1. Temperature is obtained using the latest calibration from Alonso et al.[14].

2. Surface gravity is obtained using parallax taken from Hipparcos catalog[15].

3. Age calculation is based on the latest evolutionary tracks and isochrones provided by VandenBerg et al.[16]

With these improvements, the calculations for abundances and ages of the samples are more reliable. More importantly, this data is completely consistent with that of Chen et al.[9], which is a basic precondition in obtaining reliable results in large sample statistical analysis. An important point to note is that in regard to the 25 common samples used by Chen et al.[9] and Edvardsson et al.[10], we adopt only the data of Chen et al.[9] considering that the results of the former are of higher accuracy. In this paper, abundances of 10 elements are analyzed, with the emphasis on Fe.

Orbital parameters used in this paper are calculated with the program developed by Cui Chenzhou, which is also used in Cui2000. The galactic mass distribution model provided by Allen and Santillán [17] is adopted in the program as a potential function. According to this galactic mass distribution model, the Galaxy is made up of three components: a spherical central bulge, a disk and a massive spherical halo. The total mass of the Galaxy resulting from the model is $9\times10^{11}M_{\odot}$. In terms of mathematics, the model is fully analytical, continuous, and has continuous derivatives everywhere. Results from the model are coherent with the observed rotation curve and perpendicular force $F_z$ at the Sun location. Adopted by the program, the galactocentric distance of the Sun ($R_0$) is 8.5 kpc, the circular velocity of the Sun is 220 km $s^{-1}$[18], and the space velocities ($U$, $V$, $W$) of the Sun are (−10, 5.2, 7.2) km $s^{-1}$ [19].

In order to get reliable statistical results, samples with large error are eliminated. Additionally, it is necessary to eliminate halo stars to ensure that all of the samples are disk stars. Following these criteria, we eliminate 19 samples shown as follows:

1. HD5750. Orbital parameters cannot be calculated since no parallax data can be found in the Hipparcos catalog.

2. HD30649A, HD62301, HD201099A and HD201891. Orbital parameters cannot be calculated since there are no full sets of space velocities for them that are reliable.

3. HD54717, HD58855, HD59380, HD106516A, HD142860A, HD157089, HD157466, HD177565 and HD204363. Proportions among the samples' $U_{LSR}$, $V_{LSR}$ and $W_{LSR}$ are very abnormal, far from those of others, orbital integration error obtained from Cui2000 program is very large.

4. HD51929 and HD63077. Their mean orbital radii ($R_m$) are both greater than 18 kpc.

5. HD25704, HD97916 and HD148816. Their maximum vertical distance to the galactic disk ($Z_{max}$) is greater than 3.5 kpc.

Samples mentioned in 1, 2 and 3 must be eliminated because we cannot obtain reliable orbital parameters for them. Samples mentioned in 4 and 5 belong to different populations, and thus they must be eliminated too.

Because the origin and evolution history of the thin disk and the thick disk may be different, mixture of different types of samples will reduce the reliability of the analysis results. Generally, it is difficult to accurately define the population origin of a star, whether it is in the thin disk or the thick disk. However, according to present understanding of the general characteristics of thick disk stars, we can roughly exclude thick disk stars from thin disk stars. The main goal of this paper is to analyze thin disk samples, so a comparative loose criterion for thick disk star is adopted: $-1.6<$ [Fe/H] $\leqslant -0.4$ dex, $V_{LSR} \leqslant -40$ km $s^{-1}$, Age $\geqslant$ 7 Ga[20—22]. According to the criterion, 18 samples are ascribed to the thick disk group. In order to re-check whether these samples are thick disk stars, their [$\alpha$/Fe] and $\sigma$ ($W_{LSR}$) are analyzed, where [$\alpha$/Fe] is the mean value of [Mg/Fe], [Si/Fe], [Ca/Fe] and [Ti/Fe]. The mean value of [α /Fe] is 0.18 dex, and the σ ($W_{LSR}$) is 41 km $s^{-1}$, which are all coherent with known characteristics of thick disk stars. Similar criteria have been adopted in other papers[23,24]. After such division, the number of the remaining thin disk samples is 217.

## 2 Radial abundance gradients

The samples used in this paper are all intermediate- and low-mass late-type stars; therefore their evolution histories have a large age span. If we divide them into several subgroups according to their ages, the radial abundance gradient condition of the galactic disk in different historical periods can be obtained. When we analyze [Fe/H] of the sample data, we divide the 217 thin disk stars into 4 subgroups except when analyzing them as a whole. The age spans of these four subgroups are Age≤4 Ga, 4 Ga<Age≤6 Ga, 6 Ga < Age ≤8 Ga and Age>8 Ga respectively. There are only a few samples with Age ≤ 2 Ga or Age >10 Ga, so instead of dividing them into two subgroups, we combine them into neighboring groups. Fig. 1 shows the relationship between [Fe/H] and mean orbital radius $R_m$. The thin disk stars are shown as filled circles and the thick disk stars as open circles. The line is the linear regression analysis result for the thin disk stars. The analysis results of 10 chemical elements for the thin disk stars are listed in table 1.

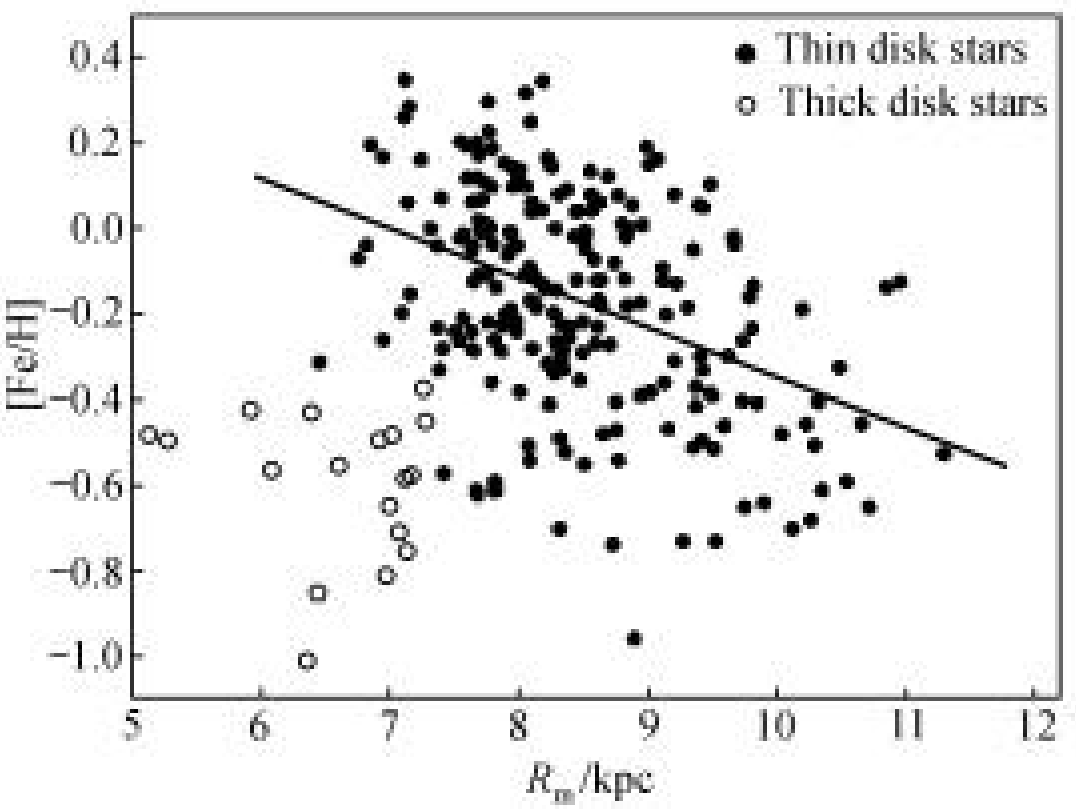

Fig. 1. Relationship between [Fe/H] and $R_m$. Thin disk stars are shown as filled circles and thick disk stars as open circles.

Table 1 Statistical results between abundances and $R_m$ a)

| Element | Division criterion | *A* | *B* | *R* | SD | *N* |
|---|---|---|---|---|---|---|
| Fe | all | 0.811±0.150 | −0.116±0.018 | 0.410 | 0.233 | 217 |
| | Age≤4 Ga | 0.666±0.205 | −0.088±0.024 | 0.378 | 0.169 | 79 |
| | 4 Ga < Age≤ 6 Ga | 0.660±0.304 | −0.096±0.036 | 0.336 | 0.219 | 56 |
| | 6 Ga<Age≤8 Ga | ~~0.263±0.399~~ | ~~−0.061±0.046~~ | ~~0.233~~ | ~~0.264~~ | ~~33~~ |
| | Age > 8 Ga | 0.938±0.301 | −0.145±0.035 | 0.521 | 0.248 | 49 |
| O | all | 0.502±0.132 | −0.065±0.016 | 0.336 | 0.165 | 140 |
| Na | all | 0.822±0.157 | −0.116±0.018 | 0.403 | 0.242 | 208 |
| Mg | all | 0.888±0.139 | −0.114±0.016 | 0.445 | 0.207 | 200 |
| Al | all | 0.740±0.170 | −0.101±0.020 | 0.337 | 0.248 | 197 |
| Si | all | 0.761±0.138 | −0.104±0.016 | 0.400 | 0.214 | 217 |
| Ca | all | 0.734±0.127 | −0.102±0.015 | 0.422 | 0.197 | 217 |
| Ti | all | 0.612±0.140 | −0.090±0.016 | 0.355 | 0.213 | 206 |
| Ni | all | 0.886±0.165 | −0.123±0.019 | 0.401 | 0.251 | 209 |
| Ba | all | 0.711±0.180 | −0.101±0.021 | 0.346 | 0.245 | 169 |

a) Regression formula: $Y = A + B \cdot X$; *A*, *B*: intercept and slope of linear regression; *R*: correlation coefficient; SD: standard deviation; *N*: sample number.

For a given number of samples, for which the correlation coefficient must be larger than a certain value, the analysis result is meaningful in terms of statistics[25]. Notice that the data on stars with ages of 6 to 8 giga years are marked with strikethrough; this is because the results of the analysis on this group does not satisfy the above statistical criterion. This group has only 33 samples; therefore the scatter is too large, which results in the deduction of the correlation coefficient. These samples have no statistical value; we keep them only for reference.

From fig. 1, we learn that the dispersion of thick disk stars is very large and an abundance gradient does not exist in the thick disk. To thin disk samples, generally, the relationship between [Fe/H] and $R_m$ can be described as

$$d[Fe/H]/dR_m = -0.116 \pm 0.018 \text{ dex kpc}^{-1}.$$

This gradient value is a little steeper than that of Cui2000. The main reason is that the orbital parameter adopted in Cui2000 is the maximum galactocentric distance $R_{max}$, while $R_{max}$ is larger than $R_m$ for both the value and the distribution of the value. Furthermore, in Cui2000, sample selection is not so strict as in this paper; the dispersion of abundance is larger, which affects the statistical result. The result of this paper is consistent with the results of open clusters given by Cameron[26] and Janes et al.[27]. An important observation is that their open cluster samples are generally young.

The gradients of the two subgroups, Age⩽4 Ga and 4 Ga<Age⩽6 Ga, are $-0.088 \pm 0.024$ dex kpc$^{-1}$ and $-0.096 \pm 0.036$ dex kpc$^{-1}$, which are consistent with results of open clusters given by Carraro et al.[28], Friel et al.[29], Friel[30] and Thogersen et al.[31]. The ages of most open clusters are less than 6 Ga, so we use the statistical results of the two subgroups of young stars for comparison.

From table 1, we learn that [Fe/H] gradients have a tendency to become steeper with increasing age; the gradients of the three subgroups with meaningful results are –0.088, –0.096 and –0.145 dex kpc$^{-1}$. Similar conclusions, i.e. younger stars have flatter abundance gradients than older ones, are obtained by Janes et al.[27], Carraro et al.[28], and Friel[32]. This means, during the chemical evolution of the Galaxy, the in-fall and out-flow processes of gas are of important influence to abundance gradients, although stellar nucleosynthesis is the dominating factor.

The statistical results of abundance gradients for other elements are also listed in table 1. O is a very important element in astrophysics. From the listed results, the gradient of O is $-0.065 \pm 0.016$ dex kpc$^{-1}$, which is consistent with the results of B-type stars given by Smartt and Rolleston[7] and Rolleston et al.[8], and the results of H II region given by Shaver et al.[33]. In α elements, as a result of similar enrichment histories, the gradients of Si and Ca are very similar. But, the gradient of Mg is more similar to those of Si and Ca than to that of O, which means that there are differences between the enrichment processes of Mg and O. Ti may not lie in a smooth extension of Si and Ca; its gradient is flatter and data dispersion is larger.

The enrichment situation for the odd-*Z* elements is more complicated. Among these, the Al gradient is flatter than that for Na. Furthermore, among all the 10 elements, the correlation coefficient between [Al/H] and $R_m$ is the faintest, and the data dispersion is the largest. Ni is an iron-peak element. Its gradient behavior is similar to Fe, but not exactly the same: the gradient value is larger than Fe, but with larger dispersion. At last, Ba, the only element that is heavier than iron-peak elements, is thought to be synthesized mainly through the neutron capture s-process. The abundance gradient of Ba is –0.101 dex kpc$^{-1}$, which is flatter than that of Fe but similar to those of Si and Ca, the two α elements. We can learn from the above that nucleosynthesis of chemical elements is an extremely complex process, and that each element seems to have a unique behavior as a result of different origin and enrichment history.

## 3 Vertical abundance gradients

Compared to radial abundance gradients, abundance gradients in vertical direction of the galactic disk are studied very little. Whether there are abundance gradients in the vertical direction has been a vexing problem for many years. Piatti et al.[34] obtain vertical gradient of $-0.34 \pm 0.03$ dex kpc$^{-1}$ within the range of 1.5 kpc, based on the study of [Fe/H] of 233 red giants in 63 open clusters. The vertical gradient given by Cui2000 is $-0.411 \pm 0.023$ dex kpc$^{-1}$. However, no vertical gradient is found in studies of Friel et al.[29], Friel[30] and Cameron[26]. In the case of field stars, orbital dispersion can often smear the vertical gradient of single age samples. The statistical results of vertical gradients are more uncertain than radial gradients.

After analysis of the abundance gradients of the galactic disk in the radial direction, in the following we will briefly analyze the abundance gradients for the vertical direction. Considering that the vertical height for thin disk stars is usually less than 0.8 kpc, and that there are only 17 samples with maximum vertical distance $Z_{max} > 0.8$ kpc in the whole sample (less than 5% of the whole), and that furthermore these few marginal data points will enlarge the error of the statistical results, we will only analyze the samples with $Z_{max} < 0.8$ kpc. The statistical results between [Fe/H] and $Z_{max}$ are shown in fig. 2, where thin disk stars are shown as filled circles, thick disk stars as open circles and merged data as open squares. The span of the merged thin disk data is 0.1 kpc. The last merged data is the result of 4 samples with $Z_{max} > 0.6$ kpc.

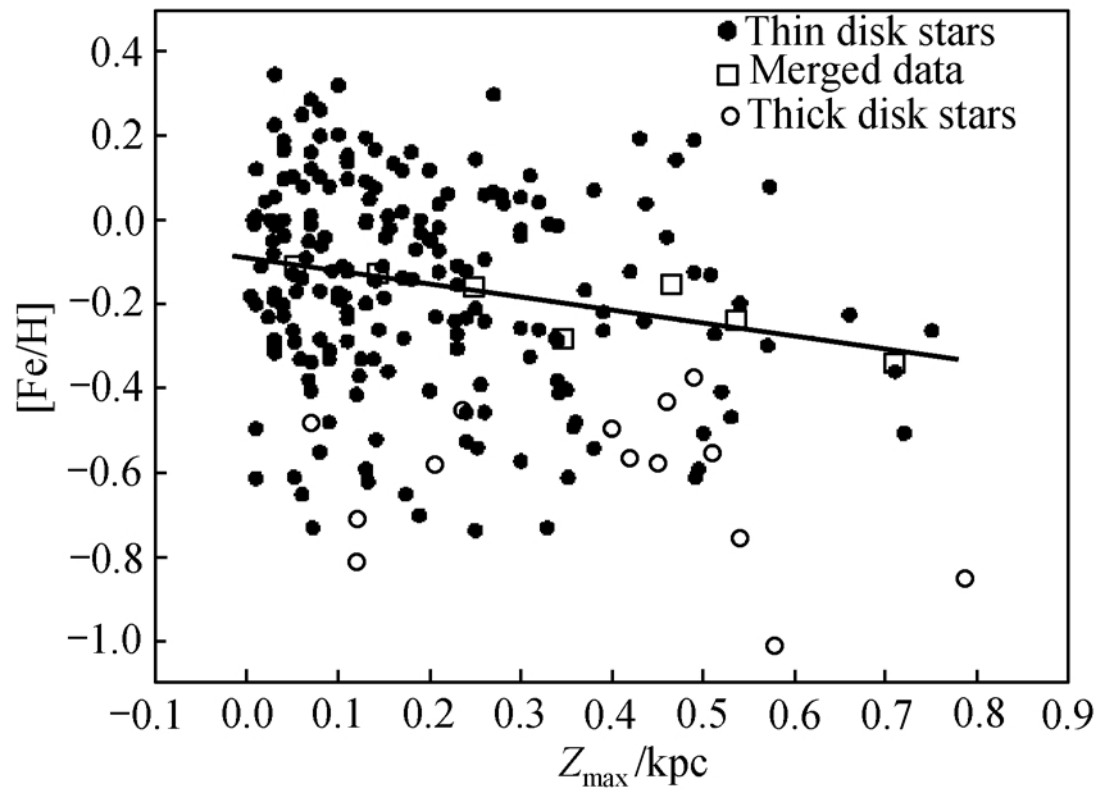

Fig. 2. Relationship between [Fe/H] and $Z_{max}$. Thin disk stars are shown as filled circles, thick disk stars as open circles and merged data as open squares.

In fig. 2, no gradient can be seen in the thick disk samples. Although the dispersion is very large, the mean [Fe/H] of the thin disk samples with $Z_{max} < 0.6$ kpc is in general higher than that of samples with $Z_{max} > 0.6$ kpc. We merge these data so that the general tendency of abundances can be seen. Correlation is obvious among the 7 merged data groups. The gradient is d[Fe/H]/d$Z_{max}$ = – 0.309±0.098 dex kpc$^{-1}$, not far from the result of Cui2000 and consistent with Piatti et al.[34]. Furthermore, we notice that the merged data of the samples whose $Z_{max}$ is between 0.3 kpc and 0.4 kpc is lower than the general tendency obtained from others. A similar situation can be found in fig. 2 of Piatti et al.[34]. The reason why the merged data is lower than the general tendency at this position is unknown at present.

An ideal method for studying the distribution of abundances as a function of $Z_{max}$ is to divide samples into groups according to their age and then sub-divide each group into smaller groups according to their mean orbital radius. Then the abundance distribution in a certain historical period at a certain place in the galactic disk can be obtained. Limited by the number of samples, we cannot carry out such a two-level division according to the above criteria. Otherwise the sample number of each group will be too small from which to obtain meaningful statistical results. The analysis results between [Fe/H] and $Z_{max}$ of thin disk stars are listed in table 2, including results of whole samples, merged data and sub-grouped data. Strikethroughs in this table have the same meanings as in table 1.

Table 2 Statistical results between [Fe/H] and $Z_{max}$ a)

| Division criterion | *A* | *B* | *R* | SD | *N* |
|---|---|---|---|---|---|
| All | –0.088±0.027 | –0.326±0.106 | 0.213 | 0.237 | 200 |
| Merged | –0.090±0.041 | –0.309±0.098 | 0.816 | 0.055 | 7 |
| Age≤4 Ga | –0.008±0.030 | –0.320±0.139 | 0.257 | 0.173 | 77 |
| 4 Ga<Age≤6 Ga | ~~–0.097±0.051~~ | ~~–0.219±0.221~~ | ~~0.137~~ | ~~0.225~~ | ~~53~~ |
| 6 Ga<Age≤8 Ga | ~~–0.328±0.089~~ | ~~0.272±0.309~~ | ~~0.161~~ | ~~0.275~~ | ~~31~~ |
| Age>8 Ga | –0.160±0.065 | –0.444±0.221 | 0.313 | 0.267 | 39 |
| $R_m$≤8 kpc | ~~–0.043±0.043~~ | ~~–0.194±0.200~~ | ~~0.113~~ | ~~0.229~~ | ~~74~~ |
| 8 kpc<$R_m$≤ 9 kpc | –0.076±0.036 | –0.322±0.155 | 0.227 | 0.214 | 82 |
| $R_m$ > 9 kpc | ~~–0.283±0.061~~ | ~~–0.106±0.192~~ | ~~0.085~~ | ~~0.240~~ | ~~44~~ |

a) All factors have the same meanings with table 1.

Based on the meaningful statistical data listed in table 2, all the values of d[Fe/H]/d$Z_{max}$ are between – 0.3 dex kpc$^{-1}$ and – 0.45 dex kpc$^{-1}$, not far from the result of all the samples without being sub-grouped. Additionally, a similar tendency as in section 2 (the flattening of abundance gradients as the Galaxy evolves) can be obtained from the two meaningful results of the samples that are sub-grouped according to age.

## 4 Discussion and conclusions

In this paper, we analyze 235 F and G-type stars which have accurate abundances and kinematics. The variance of chemical abundance gradients in the galactic disk on a time and space scale is investigated by dividing samples into different subgroups according to their age and mean orbital radius. The main results are shown as follows:

1) In the thin disk, there are obvious abundance gradients. Gradient values are consistent with

the results obtained by other researchers. The gradient tends to become flatter with the evolution of the Galaxy, which means that in-fall and out-flow processes play important roles in the distribution of abundances.

2) In the thick disk, based on the only 18 samples, no gradient is found in both radial and vertical directions.

3) Statistical results of this paper indicate that the ELS model is mainly suitable for the evolution of the thin disk, while the SZ model is more suitable for the evolution of the thick disk.

It should be emphasized that dispersions in all the statistical results are quite large, not only in this paper, but also in the papers of other researchers. One possible reason is that large dispersion is the result of a small number of samples, bias of sample selection and observation, calculation error. For example, in fig. 1, we can hardly find correlation between [Fe/H] and $R_m$ if we include the thick disk samples in analysis. This means a mixture of different populations will greatly change the statistical results, which should be considered carefully when carrying out work in this field.

But the greater possibility may be that the large dispersion of abundance gradients is the result of the intrinsic complexity of the galactic evolution. In-fall, out-flow, radial flow, interaction among stellar objects, and so on, will enlarge the dispersion. Quantitative analysis with much higher resolution and much greater precision will improve the statistical precision and the reliability to some degree, but may not diminish the dispersion. A reasonable suggestion is that we should keep in mind the fact that the dispersion of the abundance gradients is large. Meanwhile we must devote more attention to building more reasonable galactic chemical evolution models according to this fact.

**Acknowledgements** This research was supported by the National Natural Science Foundation of China (Grant Nos. 10173014 and NKBRSF G1999075406).